# A Novel Cluster Classify Regress Model Predictive Controller Formulation ; CCR-MPC

Clement Etienam[1*] , Siying Shen[1] , Edward J O'Dwyer[2] & Joshua Sykes[1]

[1] *Active Building Research Programme*

[2] *Imperial College London*

*Corresponding Author*

# Abstract

In this work, we develop a novel data driven model predictive controller using advanced techniques in the field of machine learning. The objective is to regulate control signals to adjust the desired internal room set point temperature , affected indirectly by the external weather states. The methodology involves developing a time-series machine learning model with either a Long Short Term Memory model (LSTM) or a Gradient Boosting Algorithm (XGboost), capable of forecasting this weather states for any desired time horizon and concurrently optimising the control signals to the desired set point. The supervised learning model for mapping the weather states together with the control signals to the room temperature is constructed using a previously developed methodology called Cluster Classify regress (CCR), which is similar in style but scales better to high dimensional dataset than the well-known Mixture-of-Experts. The overall methodology involves using the CCR as a forward model in a batch or sequential optimisation approach. The overall method called CCR-MPC involves a combination of a time series model for weather states prediction, CCR for forwarding and any numerical optimisation method for solving the inverse problem . Two numerical method for optimisation are shown, Nelder–Mead Approximation and the Bayesian style, iterative ensemble smoother(I-ES). Forward uncertainty quantification (Forward-UQ) leans towards the regression model in the CCR and is attainable using a Bayesian deep neural network or a Gaussian process (GP). For this work, in the CCR modulation, we employ K-means clustering for Clustering , XGboost classifier for Classification and $5^{th}$ order polynomial regression for Regression. Inverse UQ can also be obtained by using an I-ES approach for solving the inverse problem or even the well-known Markov chain Monte Carlo (MCMC) approach. The developed CCR-MPC is elegant, and as seen on the numerical experiments is able to optimise the controller to attain the desired set point temperature.

# 1. Introduction

Energy efficiency is a major concern to achieve sustainability in modern society. Smart cities sustainability depends on the availability of energy-efficient infrastructures and services. Buildings compose most of the city, and they are responsible for most of the energy consumption and emissions to the atmosphere (40%). Smart cities need smart buildings to achieve sustainability goals. Building's thermal modelling is essential to face the energy efficiency race. In this report, we introduce a novel data driven model predictive controller (MPC). A mathematical introduction is presented to help understand the fundamental supervised learning problem to predict the room temperature in a building and also forecast weather states for 10-15 minutes lag duration. Finally the overall data driven MPC is implemented on some toy problems.

The rest of the report is structured in this manner, section 2 introduces the necessary background necessary for our novel data driven MPC methodology. This session consists of the supervised learning theory, time series modelling using recurrent neural network (RNN) and Long short term memory network (LSTM), optimisation methods, neural networks, gradient boosting and a novel cluster classify regress algorithm developed in [6] and reformulated [8]. Section 3 & 4 shows some numerical experiments and section 5 gives conclusions and insights into future work.

## 2  Background

### 2.1 Forward problem- Supervised Learning

For a supervised learning mode the following ansatz holds;

Assume $\{(x_i, y_i)\}_{i=1}^{N}$ where $x_i \in \mathbb{R}^K$ and $y_i \in \mathbb{R}^M$ for regression or $y_i \in \{0,1,...J\}^M$ for classification, where $(x_i, y_i) \in \chi \times \gamma$ assumed to be input and output of a model, Postulating amongst a family of ansatz $f(.;\theta)$, parametrized by $\theta \in \mathbb{R}^P$, we can find a $\theta^*$ such that for all $i = 1,...N$,

$$y_i = f(x_i; \theta^*)$$

Eqn.1(a)

$f: \chi \to \gamma$ is the forward mapping irregular and has sharp features, very non-linear and has noticeable discontinuities,

Then $\forall\ x'$ in the set $\{x_i\}_{i=1}^{N}$,

$$y' \approx f(x'; \theta^*)$$

<div style="text-align: right">Eqn 1(b)</div>

Where $y'$ is the true label of $x'$

### 2.1.1 Parametric models
Example includes; generalized linear models (GLM), Neural networks (NN). The idea is that;

- $N$ data points only seen in training and often not all at once, memory is sublinear or constant in $N$

### 2.1.2 Non-Parametric Models
- $N$ data points and $N$ inner product, that is required for prediction
- Bayesian
- Number of hyper-parameters $P$ can be small (3 for Gaussian processes [5])
- Seeks to identify the best model $f(.; X, Y, \theta)$ without relying on any parametric form.

### 2.1.3 Classical (Gradient) approach for model optimisation:
Find best $\theta^*$ which minimizes data misfit;

$$\Phi(\theta) = \sum_{i=1}^{N} d\big(y_i, f(x_i; \theta)\big)$$

<div style="text-align: right">Eqn 1(c)</div>

$$\theta^* = (X^T X)^{-1} X^T Y$$

<div style="text-align: right">Eqn 1(d)</div>

Where $X = [x_1, \ldots, x_N]$ and $Y = [y_1, \ldots, y_N]$

## 2.2 . Recurrent neural network (RNN)  & Long Short term memory(LSTM) Theory

An RNN model captures a time series sequence from the following ansatz; [15 &19]

Assume a hidden state $h_t \in \mathcal{H} \subset \mathbb{R}^{n_h}$, the dynamics are described as follows,

$$h_t = f_h(h_{t-1}, x_t)$$

<div style="text-align: right">Eqn 2</div>

$$y_t = f_y(h_t)$$

<div style="text-align: right">Eqn 3</div>

$f_h: \mathcal{H} \times X \to \mathcal{H}$ is the transition nonlinear function describing the dynamics of $h_t \in \mathcal{H}$ and $f_y : \mathcal{H} \to Y$ describes the functional mapping from the hidden state to the output variable.

$h_t \in \mathcal{H}$ is the running summary of $u_t \in U$ until time $t \in \mathbb{N}$. The reoccurrence formula updates this summary based on its previous value, $h_{t-1} \in \mathcal{H}$ where we assume $h_0 = 0$

$f_h$ is the composition of an element wise nonlinearity with an affine transformation of both $u_t$ and $h_{t-1}$ such that,

$$h_t = f_h(W_{uh}u_t + W_{hh}h_{t-1})$$

Eqn 4

$W_{uh} \in \mathbb{R}^{n_h \times n_u}$ is the input to hidden weight matrix

$W_{hh} \in \mathbb{R}^{n_h \times n_h}$ is the hidden to hidden weight matrix

$$y_t = f_y(W_{hy}h_t)$$

Eqn 5

Where $W_{hy} \in \mathbb{R}^{n_y \times n_h}$ is the hidden to output weight matrix

Defining $f_h$ as a sigmoid or hyper-bolic tangent function and $f_y$ as an identity function, we would have,

$$h_t = tanh(W_{uh}u_t + W_{hh}h_{t-1})$$

Eqn 6

and also,

$$y_t = (W_{hy}h_t)$$

Eqn 7

Where $\tanh(x) = \frac{e^{2x}-1}{e^{2x}+1}$

In (6), $W \equiv \{W_{uh}, W_{hh}, W_{hy}\}$ are parameters that are estimated using back propagation method through time (BPTT) . RNN's have memory ability, but long short term memory is limited. This makes the gradient explode or varnish while training using BPTT algorithm.[15]

An LSTM unit consists of *memory* cell $c_t$ an *input* gate $i_t$, a *forget* gate $f_t$ and an *output* gate $o_t$. The memory cell carries the memory content of the LSTM unit while the three gates control the amount of changes to an exposure of the memory content.

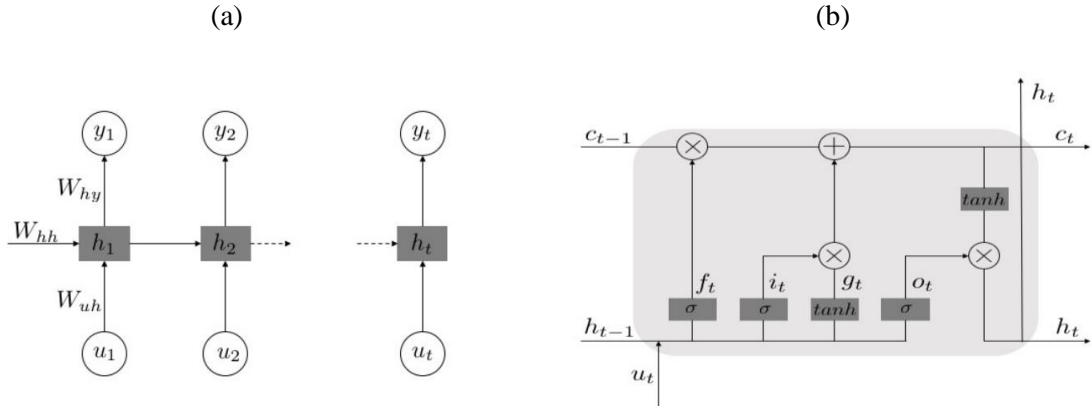

**Fig 1:** The architecture of an (a) RNN and (b) LSTM.

In Fig.1.(b), $u_t$ is the input to the memory cell layer, and $\sigma$ is the element wise logistic sigmoid function describes as $\sigma(x) = \frac{1}{1+e^{-x}}$.

Mathematically, for $n_p$ LSTM units, the forget, input and output gates are described as follows,

$$f_t = \sigma(W_{uf}u_t + W_{hf}h_{t-1})$$

Eqn 8a

$$i_t = \sigma(W_{ui}u_t + W_{hi}h_{t-1})$$

Eqn 8b

$$o_t = \sigma(W_{uo}u_t + W_{ho}h_{t-1})$$

Eqn 8c

$W_{uf} \in \mathbb{R}^{n_p \times n_u}$, $W_{ui} \in \mathbb{R}^{n_p \times n_u}$ and $W_{uo} \in \mathbb{R}^{n_p \times n_u}$ are the LSTM weights from the input $u_l$ to $f_l$, $i_l$ and $o_l$ respectively. Similarly, $W_{hf} \in \mathbb{R}^{n_p \times n_h}$, $W_{hi} \in \mathbb{R}^{n_p \times n_h}$ and $W_{ho} \in \mathbb{R}^{n_p \times n_h}$ are the weights from the hidden state $h_{t-1}$ to $f_t$, $i_t$ and $o_t$ respectively. The gates from 7a-7c control the information flow through the LSTM. The forget gate $f_t$ determines how much of the hidden state $h_{t-1}$ is allowed tom pass through; the input gate $i_t$ determines the past values of the hidden state, $h_{t-1}$ that will be updated; and the output gate $o_t$, determines how much of $h_{t-1}$ will be made available to the next layer. The final candidate value , $g_t$ and the memory cell, $c_t$ are updated by;

$$g_t = tanh(W_{ug}u_t + W_{hg}h_{t-1})$$

Eqn 9

$$c_t = f_t c_{t-1} + i_t g_t$$

Eqn 10

$$h_t = o_t \tanh(c_t)$$

Eqn 11

Finally an RNN with LSTM architecture is implemented by replacing the recurrent hidden layer with an LSTM cell.

## 2.3 Bayesian Optimisation methods (UQ)

### 2.3.1 Iterative Ensemble Smoother (I-ES)

We give a brief introduction here, for more the reader may refer to [29]

Let $y = f(x; \theta) + \varepsilon$ where $\varepsilon$ is random noise .Let $X = [x_1, \ldots, x_N]$ and $= [y_1, \ldots, y_N]$, we place a prior on $\theta$, the aim is to sample from the posterior

$$p(\theta|X,Y) \propto p(Y|X,\theta) p(X|\theta) \, p(\theta)$$

Eqn 12

A typical method would be using the ensemble Kalman filter (EnKF). Assuming a linear-Gaussian case, the prior pdf is expressed as

$$p(\theta) \propto a \exp\left(-\frac{1}{2}(\theta - \theta_{prior})^T C_\theta^{-1}(\theta - \theta_{prior})\right)$$

Eqn 13

In the equation above, $\theta_{prior}$ =Best prior estimate (mean) of hyper- parameters for the machine, $\theta$=Model hyper-parameters vectors, $C_\theta$=covariance matrix of the model ,$a$=constant.

Corrupting the true data,

$$Y_{obs} = Y_{true} + \varepsilon$$

Eqn 14

$y_{obs}$ =Observed data , $y_{true}$=true output data, $\varepsilon$ =the noise that accounts for the limited functionality of the measurement equipment. For full Gaussian linearity, we can assume the pdf of likelihood to be,

$$p(Y|X,\theta) = b \exp\left(-\frac{1}{2}(Y_{obs} - f(X;\theta))^T C_{Y_{obs}}^{-1}(Y_{obs} - f(X;\theta))\right)$$

Eqn 15

$C_{Y_{obs}}$ =covariance matrix of the measurement noise. The conditional pdf can then be derived as,

$$p(\theta | X, Y) = c \exp\left(-\frac{1}{2}((\theta - \theta_{prior})^T C_\theta^{-1}(\theta - \theta_{prior})\right)$$
$$\times \left(-\frac{1}{2}(Y_{obs} - f(X;\theta))^T C_{Y_{obs}}^{-1}(Y_{obs} - f(X;\theta))\right)$$

Eqn 16

$c$ =normalizing constant. The aim is the minimisation of the objective function Q (m) in the equation below,

$$Q(\theta) = Qm(\theta) + Qd(\theta)$$

Eqn 17

$$Q(\theta) = \frac{1}{2}(\theta - \theta_{prior})^T C_\theta^{-1}(\theta - \theta_{prior}) + \frac{1}{2}(Y_{obs} - f(X;\theta))^T C_{Y_{obs}}^{-1}(Y_{obs} - f(X;\theta))$$

Eqn 18

$Qm(\theta)$ = model mismatch term that provides normalisation for the Hessian matrix.
$Qd(\theta)$ =data mismatch term

If we assumed the relationship between model and predicted data is linear i.e. $f(X;\theta) = G\theta$, then the posterior mean in Eqn 18 is

$$\theta^{k+1} = \theta^k + C_\theta G^T (GC_\theta G^T + C_D)^{-1}(Y_{obs} - G\theta^k)$$

Eqn 19

Let $Y_{sim} = G\theta$ be the solution of the forward problem,

$$\tilde{C}_{\theta Y}^k = \frac{1}{N_e - 1}\sum_{j=1}^{N_e}(\theta_{uc,j} - \bar{\theta}^k)\left(Y_{sim_j}^k - \overline{Y_{sim}^f}\right)^T$$

$$= \frac{1}{N_e - 1}\sum_{j=1}^{N_e}(\theta_j^k - \bar{\theta}^k)\left(\bar{G}(\theta_j^k - \bar{\theta}^k)\right)^T = C_\theta G^T$$

Eqn 20(a)

$$\tilde{C}_{YY}^k = \frac{1}{N_e - 1} \sum_{j=1}^{N_e} \left(Y_{sim_j}^k - \overline{Y_{sim}^f}\right)\left(Y_{sim_j}^k - \overline{Y_{sim}^f}\right)^T$$

$$= \frac{1}{N_e - 1} \sum_{j=1}^{N_e} \left(\bar{G}(\theta_j^k - \bar{\theta}^k)\right)\left(\bar{G}(\theta_j^k - \bar{\theta}^k)\right)^T = G C_\theta G^T$$

Eqn 20(b)

$$\theta^{k+1} = \theta^k + \tilde{C}_{\theta Y}^k \left(\tilde{C}_{YY}^k + C_D\right)^{-1}(Y_{obs} - G\theta^k)$$

Eqn 21

Where $\tilde{C}_{\theta Y}^k \left(\tilde{C}_{YY}^k + C_D\right)^{-1}$ is the Kalman gain matrix, and $k$ is the iteration index . The method shown in Eqn 21 can also be posed in an online sequential learning of the model hyper-parameters for a parametric supervised learning model

## 2.4 Methods for approximating discontinuous functions using a Cluster Classify regress (CCR) formulation

This sub- section, re-introduces a recent algorithm developed in [6] and re-formulated in [8]

For a set of labelled data $\{(x_i, y_i)\}_{i=1}^N$ , where $(x_i, y_i) \in \chi \times \gamma$ assumed to be input and output of a model shown in Eqn 22

$$y_i \approx f(x_i)$$

Eqn 22

$f: \chi \to \gamma$ is irregular and has sharp features, very non-linear and has noticeable discontinuities,

The output space is taken as $\gamma = \mathbb{R}$ and $\chi = \mathbb{R}^d$

### 2.4.1 Cluster
In this stage, we seek to cluster the training input and output pairs

$\lambda: \chi \times \gamma \to \mathcal{L} := \{1, \ldots, L\}$ where the label function minimizes,

$$\Phi_{clust}(\lambda) = \sum_{l=1}^{L} \sum_{i \in S_l} \ell(x_i, y_i)$$

$$S_l = \{(x_i, y_i); \lambda(x_i, y_i) = l\}$$

Eqn 23

Eqn 24

$\ell_l$ loss function associated to cluster $l$

$$z_i = (x_i, y_i), \ell_l = |z_i - \mu_l|^2$$

Eqn 25

$$\mu_l = \frac{1}{|S_l|} \sum_{i \in S_l} z_i \text{ where } |.| \text{ denotes the Euclidean norm}$$

### 2.4.2 Classify

$l_i = \lambda(x_i, y_i)$ is an expanded training set

$$\{(x_i, y_i, l_i)\}_{i=1}^{N}$$

Eqn 26(a)

$$f_c: \chi \to \mathcal{L}$$

Eqn 26(b)

$x \in \chi$ provides an estimate $f_c : x \mapsto f(x) \in \mathcal{L}$ such that $f_c(x_i) = l_i$ for the majority of the data. Crucial for the ultimate fidelity of the prediction. $\{y_i\}$ is ignored at this phase. The classification function minimizes

$$\Phi_{clust}(f_c) = \sum_{i=1}^{N} \phi_c(l_i, f_c(x_i))$$

Eqn 26(c)

$\phi_c : \mathcal{L} \times \mathcal{L} \to \mathbb{R}_+$ is small if $f_c(x_i) = l_i$ for example we can choose $f_c(x) = argmax_{l \in \mathcal{L}} g_l(x)$ where $g_l(x) > 0, \sum_{l=1}^{\mathcal{L}} g_l(x)$ is a soft classifier and $\phi_c(l_i, f_c(x_i)) = -\log(g_l(x))$ is a cross-entropic loss.

### 2.4.3 Regress

$$f_r : \chi \times \mathcal{L} \to \gamma$$

Eqn 27(a)

For each $(x, l) \in \chi \times \mathcal{L}$ must provide an estimate $f_r : (x, l) \mapsto f_r(x, l) \in \gamma$ such that $f_r(x, f_c(x)) \approx y$ for both the training and test data. If successful a good reconstruction for

$$f : \chi \to \gamma$$

Eqn 27(b)

Where $f(.) = f_r(., f_c(.))$ the regression function can be found by minimizing

$$\Phi_r(f_r) = \sum_{i=1}^{N} \phi_r\big(y_i, f_r(x_i, f_c(x_i))\big)$$

Eqn 27(c)

Where $\phi_r : \gamma \times \gamma \to \mathbb{R}_+$ minimized when $f_r(x_i, f_c(x_i)) = y_i$ in this case can be chosen as $\phi_r\big(y, f_r(x, f_c(x))\big) = |y - f_r(x, f_c(x))|^2$. Data can be partitioned into $C_l = \{i; f_c(x_i) = l\}$ for $l = 1, \ldots L$ and then perform $L$ separate regressions done in parallel

$$\Phi_r^l(f_r(., l) = \sum_{i \in C_l} \phi_r(y_i, f_r(x_i, l))$$

Eqn 27(d)

### 2.4.4 Scaling of the Data
$x = (x^1, \ldots x^d) \in \mathbb{R}^d$ and $y \in \mathbb{R}$ hence, have $|(x, y) - (x', y')|^2 = (y - y')^2 + |x - x'|^2$

For $j = 1, \ldots d,$

$$\tilde{x}^j = \big(x^j - min_{i \in \{1,..N\}} x_i^j\big) / \big(max_{i \in \{1,..N\}} x_i^j - min_{i \in \{1,..N\}} x_i^j\big)$$

Eqn 28(a)

$$\tilde{y} = C\big(y - min_{i \in \{1,..N\}} y_i\big) / \big(max_{i \in \{1,..N\}} y_i - min_{i \in \{1,..N\}} y_i\big)$$

Eqn 28(b)

For $C > 1$ $C = 10d$ where $d = \dim(x)$ For regression $C = 1$

### 2.4.4 Bayesian Formulation
A critique of this method is that it re-uses the data in each phase. A Bayesian postulation handles this limitation elegantly well. Recall;

$$D = \{(x_i, y_i)\}_{i=1}^{N}$$

Assume parametric models for the classifier $g_l(.\,;\theta_c) = g_l(.\,;\theta_c^l)$ and the regressor $f_r(.\,,l;\theta_r^l)$ for $l = 1, \dots L$ where $\theta_c = (\theta_c^1, \dots, \theta_c^L)$ and $\theta_r = (\theta_r^1, \dots, \theta_r^L)$ and let $\theta = (\theta_c, \theta_r)$, the posterior density has the form

$$\pi(\theta, l|D) \propto \prod_{i=1}^{N} \pi(y_i|x_i, \theta_r, l)\, \pi(l|x_i, \theta_c)\pi(\theta_r)\pi(\theta_c)$$

Eqn 29(b)

$$\pi(y_i|x_i, \theta_r, l) \propto \exp\left(-\frac{1}{2}|y_i - f_r(x_i, l; \theta_r^l)|^2\right)$$

Eqn 29(c)

And

$$\pi(l|x_i, \theta_c) = g_l(x_i; \theta_c)$$

Eqn 29(d)

$$g_l(x_i; \theta_c) = \frac{\exp\left(h_l(x; \theta_c^l)\right)}{\sum_{l=1}^{L} \exp\left(h_l(x; \theta_c^l)\right)}$$

Eqn 29(e)

$h_l(x; \theta_c^l)$ are some standard parametric regressors

### 2.5 Random Forest

Random Forests (RF) [21] are part of the algorithms called decision trees. In decision trees, the goal is to create a prediction model that predicts an output by combining different input variables. In the decision tree, each node corresponds to one of the input variables; each leaf represents a value of the target variable given the values of the input variables represented by the path from the root to the leaf. Random forest algorithm trains different decision trees by using different subsets of the training data. as depicted in Fig.2.

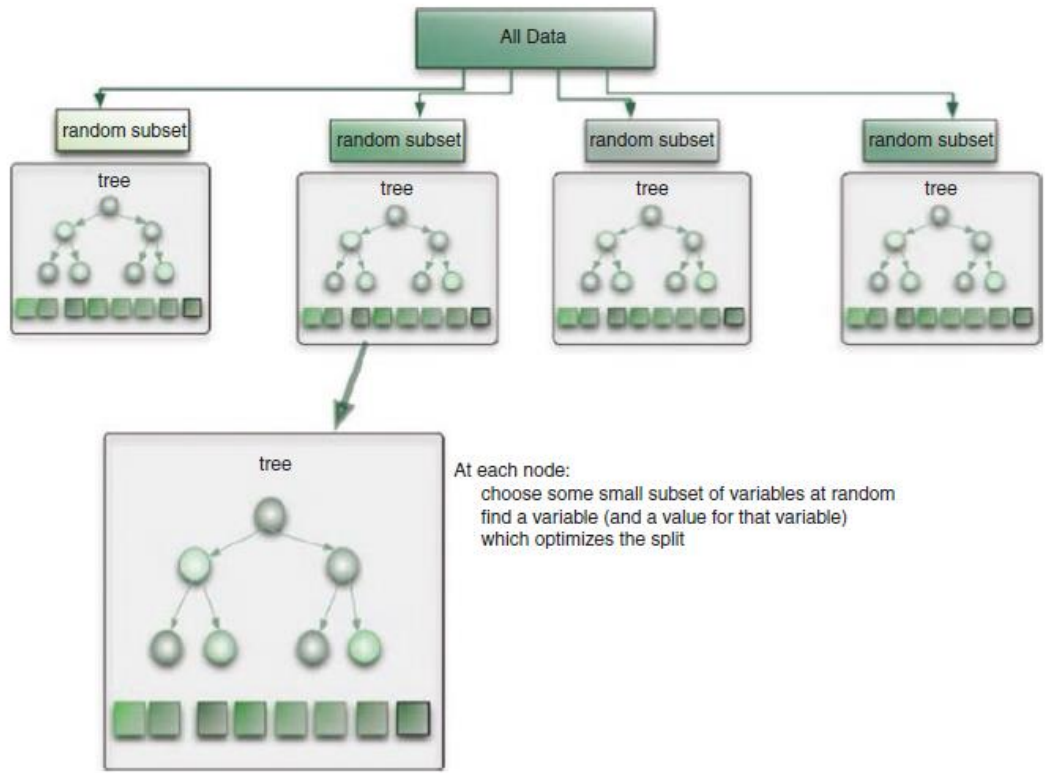

**Fig.2**: Sketch representation of a RF workflow. Random subsets of the data are trained by the RF. The randomness generates models that are not correlated to each other

### 2.5.1. Important parameters in a RF

- **Maximum features**: this is the maximum number of features that a RF is allowed to try in each individual tree.
- **Number of estimators:** This is the number of built trees before taking the maximum voting or averages of predictions.
- **Minimum sample leaf size**: The leaf is the end of a decision tree

### 2.5.2 Advantages/disadvantages of RFs
*Advantages*

- The chance of overfitting decreases, since several different decision trees are used in the learning procedure. This is very crucial for making predictions in polymer performance datasets
- RFs apply pretty well when a particular distribution of the data is not required. For example, no data normalization is needed.

- Parallelization: the training of multiple trees can be parallelized (for example through different computational slots)

*Disadvantages*

- RFs usually might suffer from smaller training datasets
- The time to train a RF algorithm might be longer compared to other algorithms.

## 2.6 Neural network

An Artificial Neural Network – ANN, with a single hidden layer can be represented graphically as follows in Fig.3.;

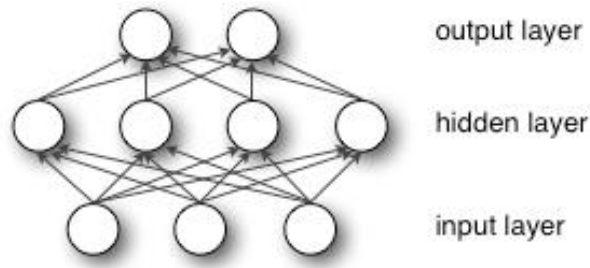

**Figure 3:** Depiction of a neural network architecture

Formally, a one-hidden-layer MLP is a function $f : \Re^K \to \Re^M$, where $K$ is the size of input vector $x$ and $M$ is the size of the output vector. We define $f(x)$, such that, in matrix notation:

$$f(x) = G\left(b^{(2)} + W^{(2)}\left(s(b^{(1)} + W^{(1)}x)\right)\right)$$

Eqn 30

With bias vectors $b^{(1)}, b^{(2)}$; weight matrices $W^{(1)}, W^{(2)}$ and activation functions $G$ and $s$

The vector $h(x) = \Phi(x) = s(b^{(1)} + W^{(1)}x)$ constitutes the hidden layer $W^{(1)} \in \Re^{K \times K_h}$ is the weight matrix connecting the input vector to the hidden layer. Each column $W_i^{(1)}$ represents the weights from the input units to the $i^{th}$ hidden unit. Typical choices for $s$ include

$$tanh(a) = \frac{(e^a - e^{-a})}{(e^a + e^{-a})}$$

Eqn 31(a)

$$sigmoid(a) = \frac{1}{(1+e^{-a})}$$

Eqn 31(b)

$$ReLU(a) = \max(0, a)$$

Eqn 31(c)

The output vector is then

$$O(x) = G\left(b^{(2)} + W^{(2)}h(x)\right)$$

Eqn 32

For classification problems, class-membership probabilities can be obtained by choosing $G$ as the *softmax* function (in the case of multi-class classification).

$$softmax\,(a)_i = \frac{exp(a_i)}{\sum_{l=1}^{L} exp(a_l)}$$

Eqn 33

Where $a_i$ represents the $i$ th element of the input to softmax which corresponds to class $i$ and $L$ is the number of classes. The result is a vector that contains the probabilities which samples $x$ that belongs to each class. The output is the class with the highest probability.

For regression the output remains

$$O(x) = \left(b^{(2)} + W^{(2)}h(x)\right)$$

Eqn 34

Where the activation function is unity. Depending on the problem type ANN uses different loss functions depending on the problem type. The loss function for classification is Cross-Entropy, which in binary case is given as,

$$Loss(\hat{y}, y, W) = -y\ln\hat{y} - (1-y)\ln(1-\hat{y}) + \frac{\alpha}{2}||W||_2^2$$

Eqn 35

where $\frac{\alpha}{2}||W||_2^2$ is an L2-regularization term (aka penalty) that penalizes complex models; and $\alpha > 0$ is a non-negative hyper parameter that controls the magnitude of the penalty. For regression problems, ANN uses the Square Error loss function; written as,

$$Loss(\hat{y}, y, W) = \frac{1}{2}||\hat{y} - y||_2^2 + \frac{\alpha}{2}||W||_2^2$$

Eqn 36

Starting from initial random weights, multi-layer perceptron (MLP) minimizes the loss function by repeatedly updating these weights. After computing the loss, a backward pass propagates it from the output layer to the previous layers, providing each weight parameter with an update value meant to decrease the loss.

In gradient descent, the gradient $\nabla Loss_W$ of the loss with respect to the weights is computed and deducted from W. More formally, this is expressed as,

$$W^{i+1} = W^i - \epsilon \nabla Loss_W^i$$

Eqn 37

where $i$ is the iteration step, and $\epsilon$ is the learning rate with a value larger than 0. The algorithm stops when it reaches a pre-set maximum number of iterations; or when the improvement in loss is below a certain, small number.

## 2.7 Gradient boosting Algorithm

We give a brief introduction here, for more information the reader may refer to [28]

For a given data set with $n$ examples and $m$ features $D = \{(x_i, y_i)\}(|D| = n, x_i \in R^m, y_i \in R)$, a tree ensemble model uses $K$ additive functions to predict the output.

$$\hat{y}_i = \varphi(x_i) = \sum_{k=1}^{K} f_k(x_i), f_k \in \boldsymbol{F}$$

Eqn 38

where $\boldsymbol{F} = \{f(x) = w_{q(x)}\}(q: R^m \to T, w \in R^T)$ is the space of regression trees ( known also as CART). $q$ represents the structure of each tree that maps an example to the corresponding leaf index. $T$ is the number of leaves in the tree. Each $f_k$ corresponds to an independent tree structure $q$ and leaf weights $w$. Each regression tree contains a continuous score on each of the leaf, $w_i$ represents score on $i$-th leaf. For a given example, we will use the decision rules in the trees (given by $q$) to classify

To learn the set of functions used in the model, we minimize the following *regularized* objective;

$$\mathcal{L}(\varphi) = \sum_i l(\hat{y}_i, y_i) + \sum_k \Omega(f_k)$$

where

$$\Omega(f) = \gamma T + \frac{1}{2}\lambda \|w\|^2$$

Eqn 39

$l$ is a differentiable *convex* loss function that measures the difference between the prediction $\hat{y}_i$ and the target $y_i$. The second term $\Omega$ penalizes the complexity of the model (i.e., the regression tree functions). The additional regularization term helps to smooth the final learnt weights to avoid over-fitting. Intuitively, the regularized objective will tend to select a model employing simple and predictive functions.

The tree ensemble model in Eq. (39) includes functions as parameters and cannot be optimized using traditional optimization methods in Euclidean space. Instead, the model is trained in an additive manner. Formally, let $\hat{y}_i^t$ be the prediction of the $i$-th instance at the $t$-th iteration, we will need to add $f_t$ to minimize the following objective.

$$\mathcal{L}^{(t)} = \sum_{i=1}^{n} l(y_i, \hat{y}_i^{(t-1)} + f_t(x_i)) + \Omega(f_t)$$

Eqn 40

This means we greedily add the $f_t$ that most improves our model according to Eq. (39). Second-order approximation can be used to quickly optimize the objective in the general setting.

$$\mathcal{L}^{(t)} \cong \sum_{i=1}^{n} [l(y_i, \hat{y}_i^{(t-1)}) + g_i f_t(x_i) + \frac{1}{2} h_i f_t^2(x_i)] + \Omega(f_t)$$

Eqn 41

where $g_i = \partial_{\hat{y}_i^{(t-1)}} l(y_i, \hat{y}_i^{(t-1)})$ and $h_i = \partial^2_{\hat{y}_i^{(t-1)}} l(y_i, \hat{y}_i^{(t-1)})$ are first and second order gradient statistics on the loss function. Removing the constant terms to obtain the following simplified objective at step $t$;

$$\tilde{\mathcal{L}}^{(t)} \cong \sum_{i=1}^{n}[g_i f_t(x_i) + \frac{1}{2} h_i f_t^2(x_i)] + \Omega(f_t)$$

Eqn 42

Define $I_j = \{i|q(x_i) = j\}$ as the instance set of leaf $j$. We can rewrite Eqn (42) by expanding $\Omega$ as follows

$$\tilde{\mathcal{L}}^{(t)} = \sum_{i=1}^{n}\left[g_i f_t(x_i) + \frac{1}{2} h_i f_t^2(x_i)\right] + \Omega(f_t) + \gamma T + \frac{1}{2}\lambda \sum_{j=1}^{T} w_j^2$$

Eqn 43

$$= \sum_{j=1}^{T}\left[\left(\sum_{i \in I_j} g_i\right) w_j + \frac{1}{2}(\sum_{i \in I_j} h_i + \lambda) w_j^2\right] + \gamma T$$

Eqn 44

For a fixed structure $q(x)$, we can compute the optimal weight $w_j^*$ of leaf $j$ by

$$w_j^* = -\frac{\sum_{i \in I_j} g_i}{\sum_{i \in I_j} h_i + \lambda}$$

Eqn 45

and calculate the corresponding optimal value by

$$\tilde{\mathcal{L}}^{(t)}(q) = -\frac{1}{2}\sum_{j=1}^{T}\frac{(\sum_{i \in I_j} g_i)^2}{\sum_{i \in I_j} h_i + \lambda} + \gamma T$$

Eqn 46

Eqn (46) can be used as a scoring function to measure the quality of a tree structure $q$. Normally it is impossible to enumerate all the possible tree structures $q$. A greedy algorithm that starts from a single leaf and iteratively adds branches to the tree is used instead. Assume that $I_L$ and $I_R$ are the instance sets of left and right nodes after the split. Letting $I = I_L \cup I_R$, then the loss reduction after the split is given by;

$$\mathcal{L}_{split} = \frac{1}{2}\left[\frac{(\sum_{i \in I_L} g_i)^2}{\sum_{i \in I_L} h_i + \lambda} + \frac{(\sum_{i \in I_R} g_i)^2}{\sum_{i \in I_R} h_i + \lambda} - \frac{(\sum_{i \in I_j} g_i)^2}{\sum_{i \in I_j} h_i + \lambda}\right] - \gamma$$

Eqn 47

Eqn (47) is used in practice for evaluating the split candidates.

## 3. Numerical Experiment 1 (Toy Problem)

**Logic of novel CCR-MPC**

- The weather state variables are $\theta$. If we assume, $\theta \in [\beta]^d$, then;

$$\theta^{t+1} = f_1(\theta^t) + \varepsilon_1$$

    Assuming a sequential approach where the weather states variable are known, the parameters of the time series machine $f_1$ can be relearned, using this logic

$$f_1^{t+1} = re-learn[f_1^t, \hat{\theta}^t]$$

    Where $t$ is time step and $\hat{\theta}^t$ is the true weather states. We then set for the next time step optimisation,

$$f_1^t = f_1^{t+1}$$

    Until final time where $f_1$ is the time series machine

- The controller is $\mu \in [\alpha]^d$ that has a direct consequence to maintain the indoor set point temperature;

$$\mu = \sum_{i=1}^{d} \alpha_i, \ldots \forall_i = 1 \ldots d$$

**d** is the dimension of the control parameters

- Hence the current weather states $\theta$ and the controller $\mu \in [\alpha]^d$ are inputs to the forward problem;

$$\gamma = f_2(\theta, \mu) + \varepsilon_2$$

Where $\gamma$ is the True temperature and $G$ is the set point temperature. $f_2$ is the supervised learning model learned from either CCR/DNN/RF, The equation (3) is then posed as an optimisation problem to track the setpoint

$$p(\mu|G, \gamma, \theta) = \underset{\mu}{\mathrm{argmin}} \|G - \gamma\|_2^2 + constraints$$

Where "constraints" could be any economical or realistic building physics scenarios we impose on the optimisation problem.

The link to GitHub Repository for implementation of the code can be found at :

https://github.com/clementetienam/Data_Driven_MPC_Controller_Using_CCR

### 3.1- Supervised Learning

A neural network (DNN) was trained for mapping the weather states with control variables to the zone mean temperature. We compute our $R^2$ accuracy for both the model learning and set point tracking by

$$R^2 = 1 - \frac{\sum_{i=1}^{N} \left[y^i_{(true\ data\ or\ setpoint)} - y^i_{(model\ prediction)}\right]^2}{\sum_{i=1}^{N} \left[y^i_{(true\ data\ or\ setpoint)} - mean(y_{(true\ data\ or\ setpoint)})\right]^2}$$

Eqn 48

Model discomfort is then

$$Discomfort(C) = \frac{1}{N} \sum_{i=1}^{N} \left[y^i_{(true\ data\ or\ setpoint)} - y^i_{(model\ prediction)}\right]^2$$

Eqn 49

In this experiment, the Nelder–Mead optimisation method [24] and the iterative ensemble smoother (I-ES) [29] are compared and used as the optimisation method.

### 3.1.1- Non Excited Room temperature machine
**Box Configuration**

*Inputs:*

- Environment: Site Outdoor Air Drybulb Temperature [C]
- Environment: Site Outdoor Air Wetbulb Temperature [C]
- Environment: Site Outdoor Air Relative Humidity [%]
- Environment: Site Wind Speed [m/s]
- Environment: Site Wind Direction [deg]
- Environment: Site Horizontal Infrared Radiation Rate per Area [W/m2]
- Environment: Site Diffuse Solar Radiation Rate per Area [W/m2]
- Environment: Site Direct Solar Radiation Rate per Area [W/m2]
- THERMAL ZONE: BOX: Zone Outdoor Air Wind Speed [m/s]

*Outputs:*

- THERMAL ZONE: BOX: Zone Mean Air Temperature [C]

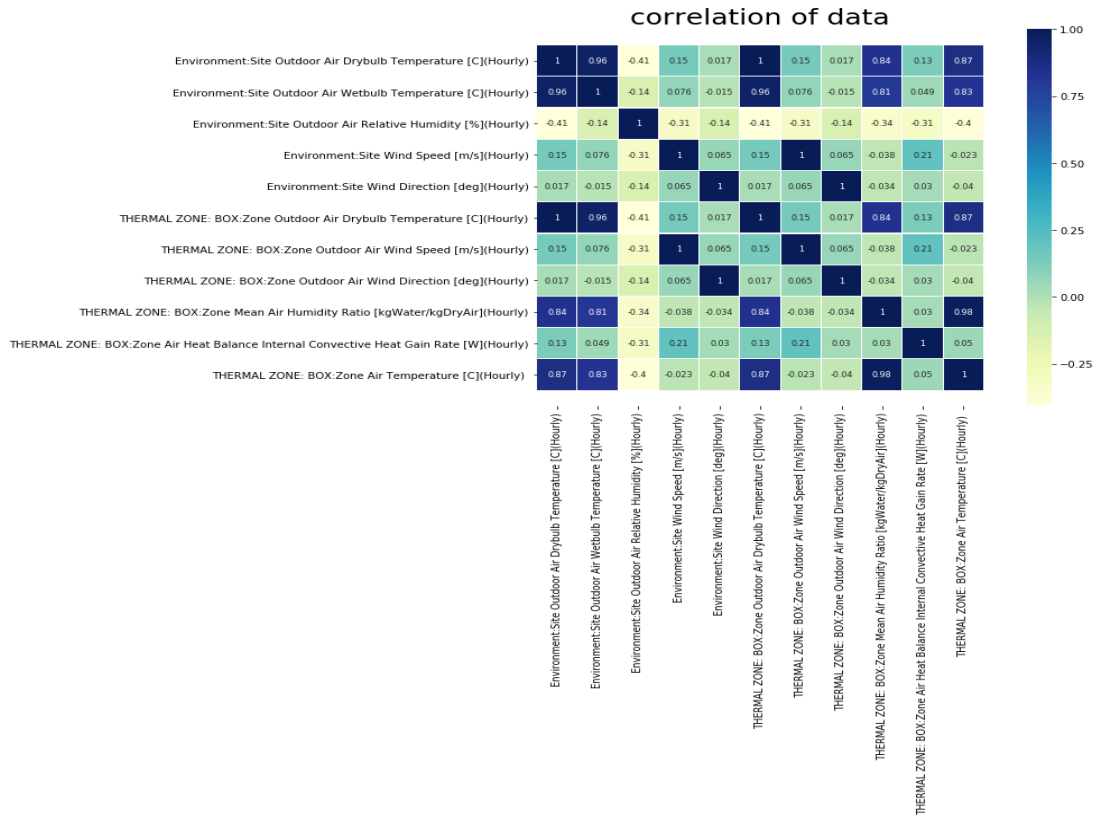

**Fig. 4:** Correlation map of the features to the outputs for Machine 1

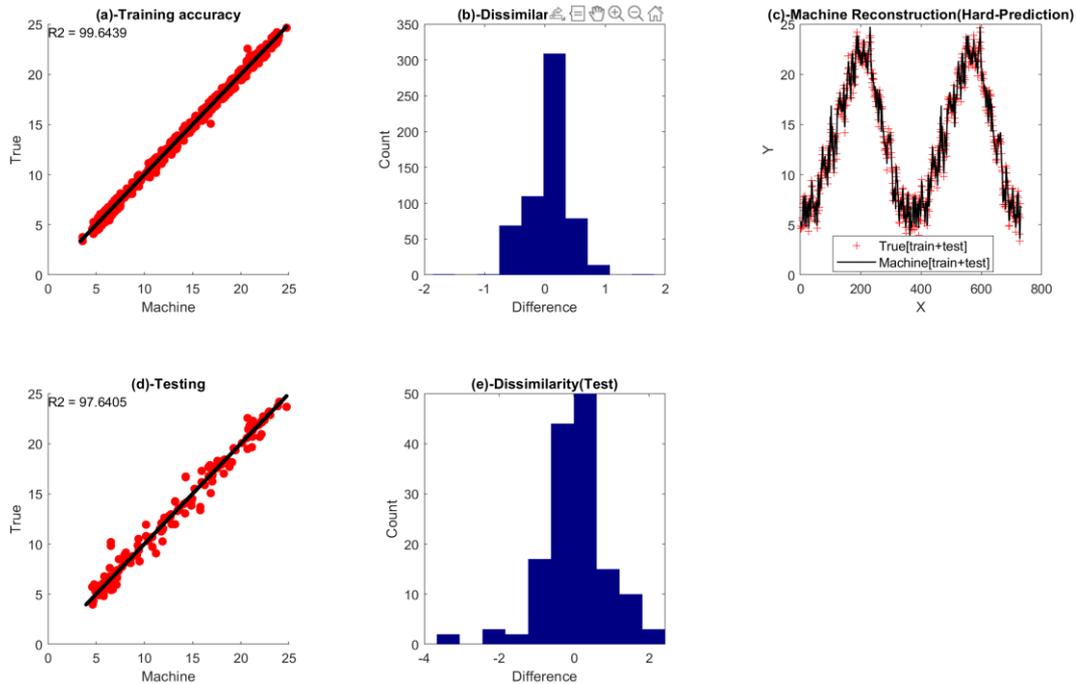

**Fig.5 :** **(a, b)-** $R^2(\%)$ training and test accuracy on Air Temperature [C](Hourly). **(b, e)-** histogram plot showing $f_1(x_i;\theta^*) - y_{true}$. **(c)-** Super imposed plot of $f_1(x_i;\theta^*), y_{true}$.

### 3.1.2- Weather states with Controller machine
**Ground Source Heat Pump (GSHP) configuration**

*Inputs:*

- Environment: Site Outdoor Air Drybulb Temperature [C]
- Environment: Site Outdoor Air Wetbulb Temperature [C]
- Environment: Site Outdoor Air Relative Humidity [%]
- Environment: Site Wind Speed [m/s]
- Environment: Site Wind Direction [deg]
- Environment: Site Horizontal Infrared Radiation Rate per Area [W/m2]
- Environment: Site Diffuse Solar Radiation Rate per Area [W/m2]
- Environment: Site Direct Solar Radiation Rate per Area [W/m2]
- THERMAL ZONE: BOX: Zone Outdoor Air Wind Speed [m/s]
- **GSHPCLG: Heat Pump Electric Power [W]-control signal**
- **GSHPCLG: Heat Pump Source Side Inlet Temperature [C]-control signal**
- **GSHPHEATING: Heat Pump Electric Power [W]-control signal**
- **GSHPHEATING: Heat Pump Source Side Inlet Temperature [C]-control signal**

*Outputs:*

- THERMAL ZONE: BOX: Zone Mean Air Temperature [C]

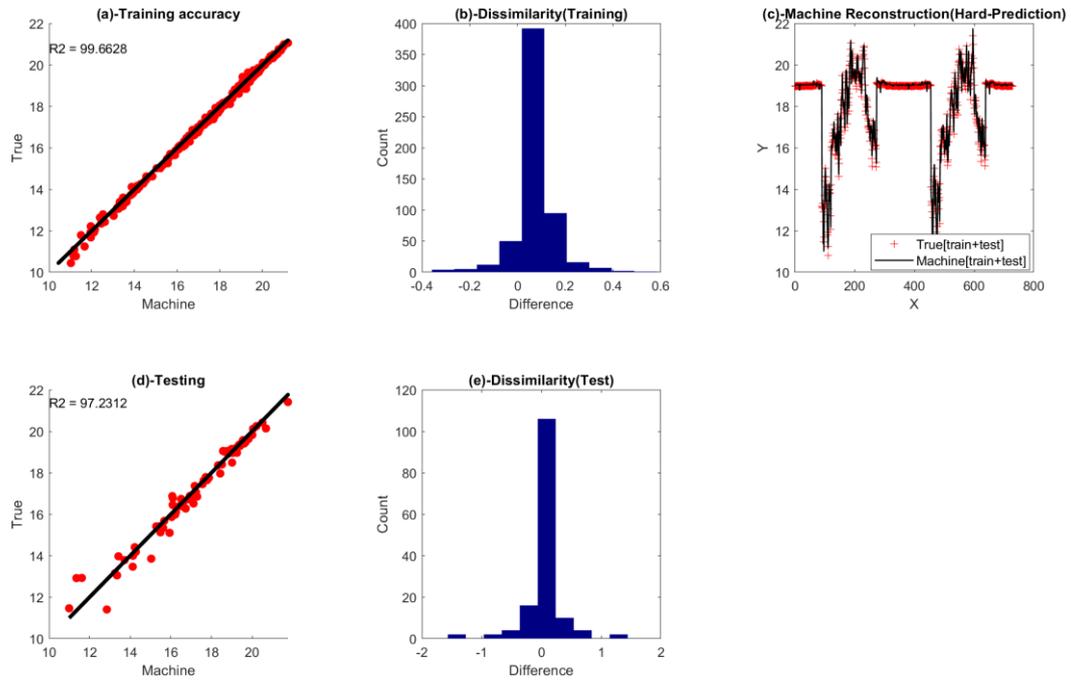

**Fig.6 : (a, b)-** $R^2(\%)$ training and test accuracy on Air Temperature [C](Hourly). **(b, e)-** histogram plot showing $f_2(x_i; \theta^*) - y_{true}$. **(c)-** Super imposed plot of $f_2(x_i; \theta^*), y_{true}$.

## 3.2 LSTM Weather states Time series modelling

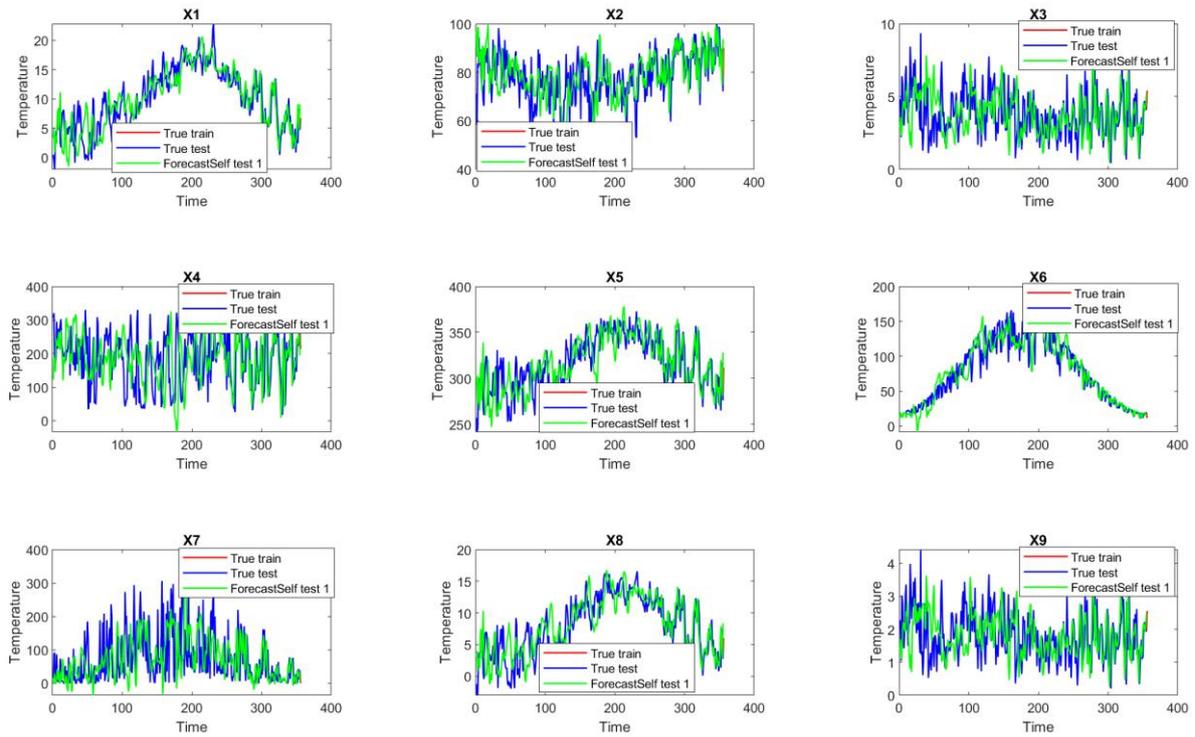

**Fig.7:** LSTM performance on the hourly prediction of the Box dataset. Each of $\mathbb{R}^9$ has been learned and forecasted to the future using only its output at its previous time step. The lookback is 7 time step, i.e $t, ..t-1.t-2, ..... t-6$ used to predict $t+1$.

## 3.3 Controller Optimisation

### 3.3.1 Batch Optimisation

(a) (b)

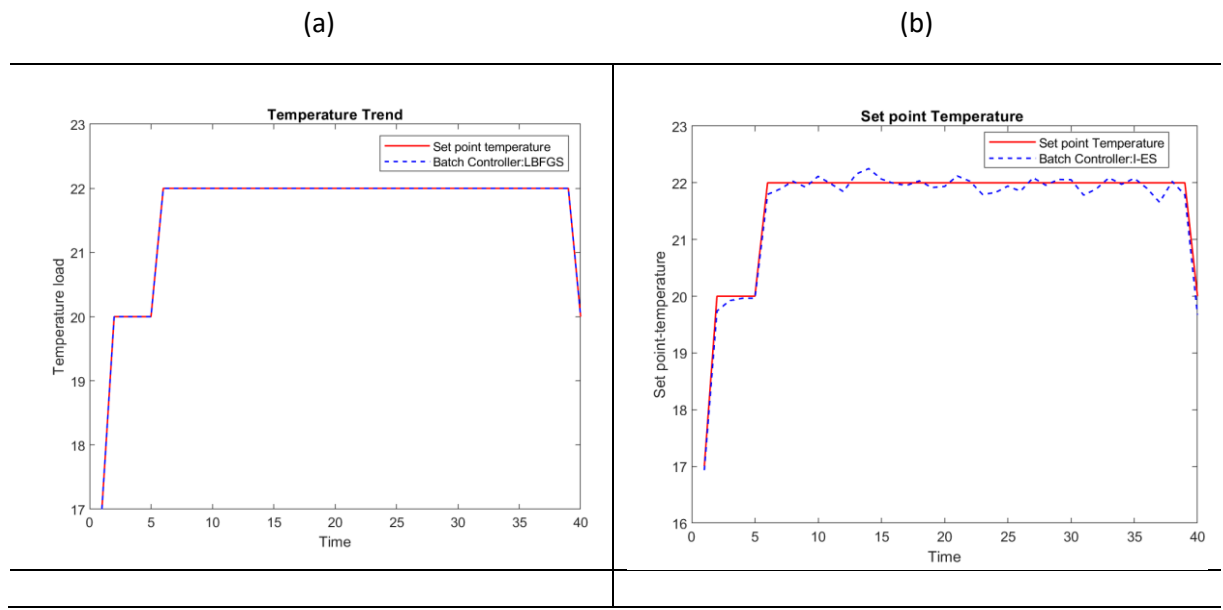

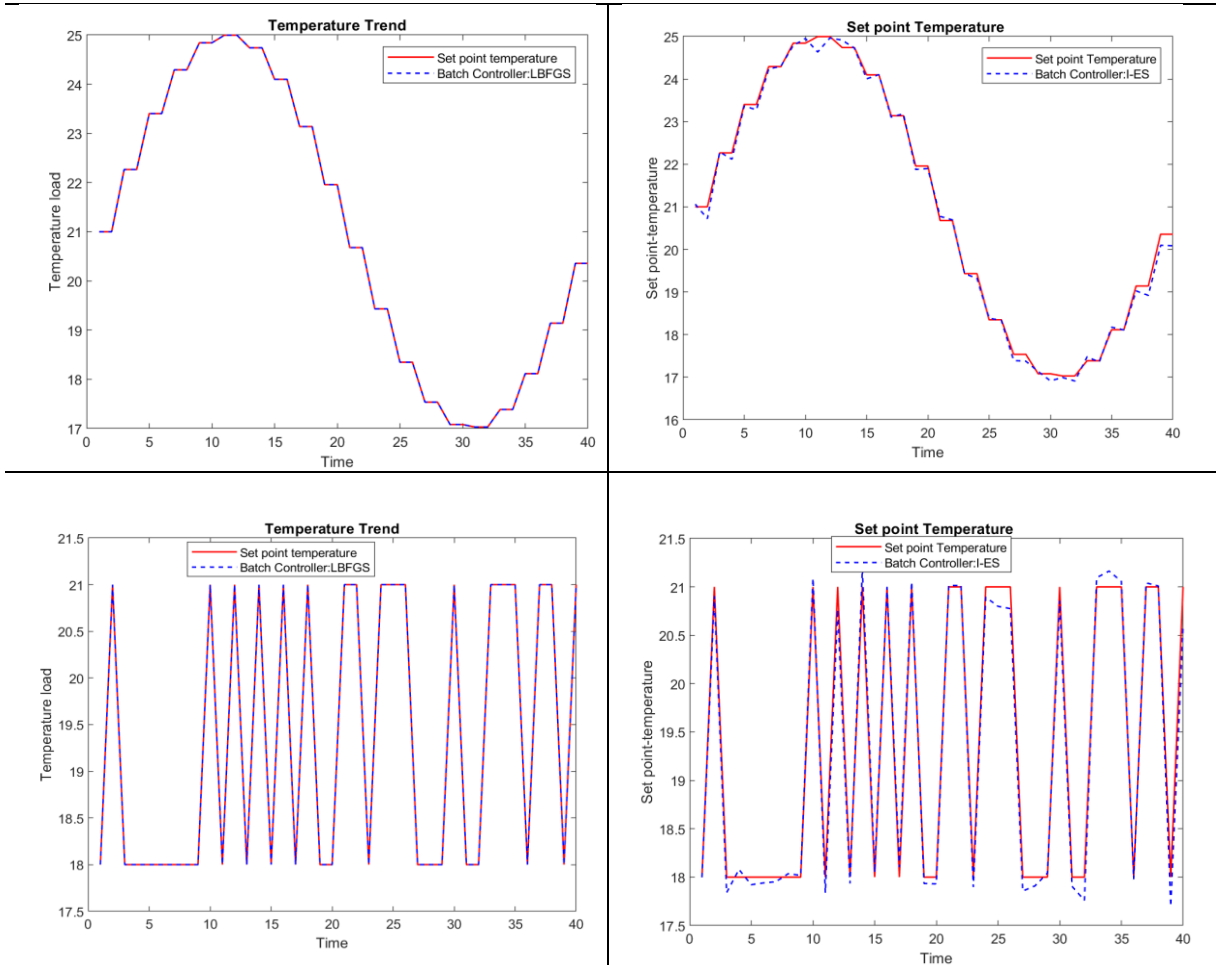

**Fig.8:** Open loop Batch Optimisation. The room temperature is forecasted "blindly" with the LSTM model, and the GSHP control parameters are optimised to the set temperature. The optimisation algorithm is Nelder–Mead on panel (a) and I-ES on panel (b)

### 3.3.2 Sequential optimisation

(a) (b)

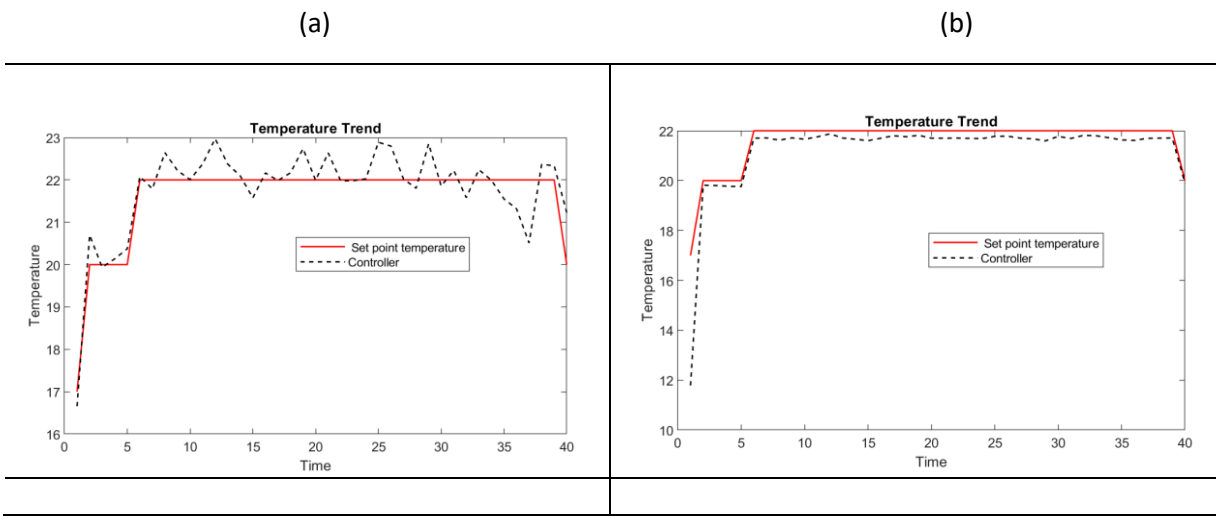

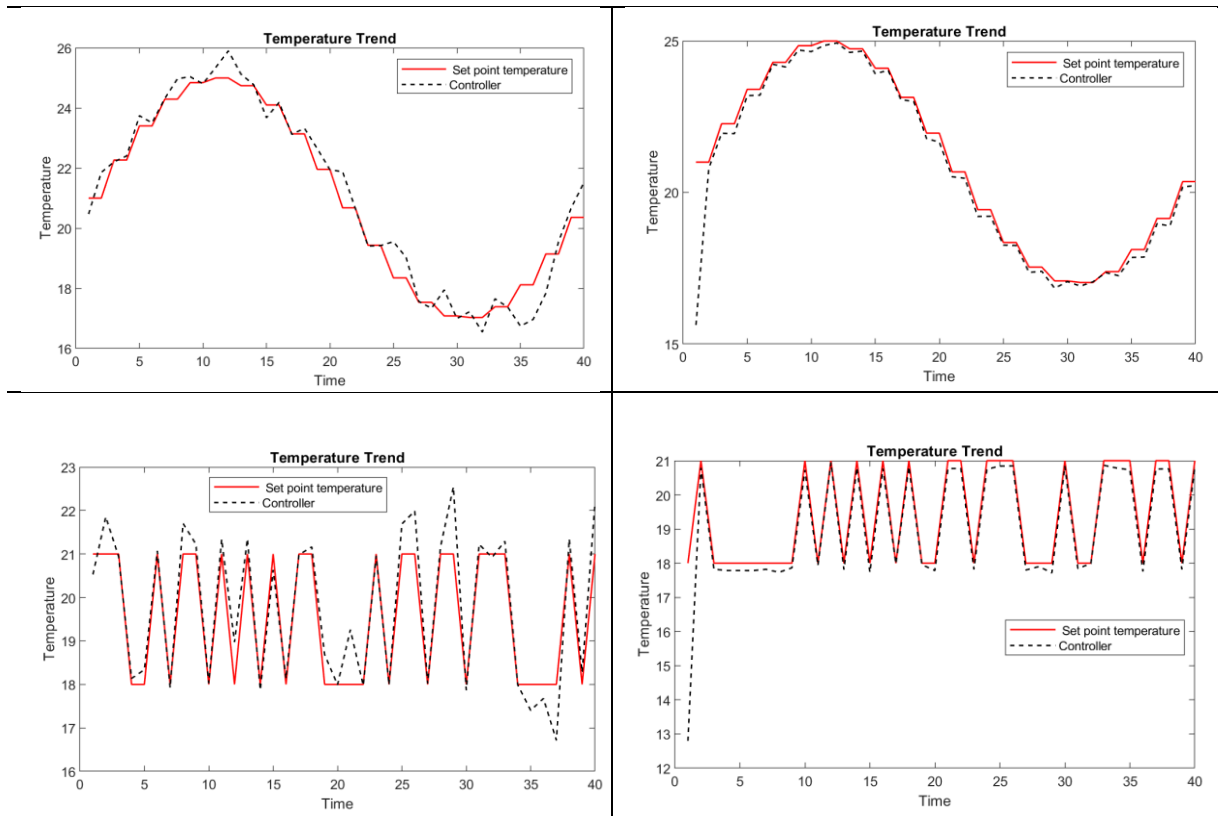

**Fig.9:** Closed loop sequential Optimisation. The room temperature is initially forecasted "blindly" with the LSTM model and corrected to the true temperature reading (from a sensor)., and the GSHP control parameters are optimised to the set temperature . The optimisation algorithm is Nelder–Mead on panel (a) and I-ES on panel (b)

## 4 Numerical Experiment 2 ( Imperial College Data)

The first step is developing a good time series model that could predict the states ( of 5 variables) accurately. Fig.10. below shows the approach using 50% of the data set and training set and 50% as test set. The time series model is the XGboost. From the "date-time" index in pandas that has this format, "01/01/2010 00:00:00", 8 new features (Pseudo-inputs), namely;

'hour', 'day of week', 'quarter', 'month', 'year', 'day of year', 'day of month', 'week of year' where created. 5 overall time series machine was modelled, mapping this 8 features to each of the 5 weather states variables.

**(a)**

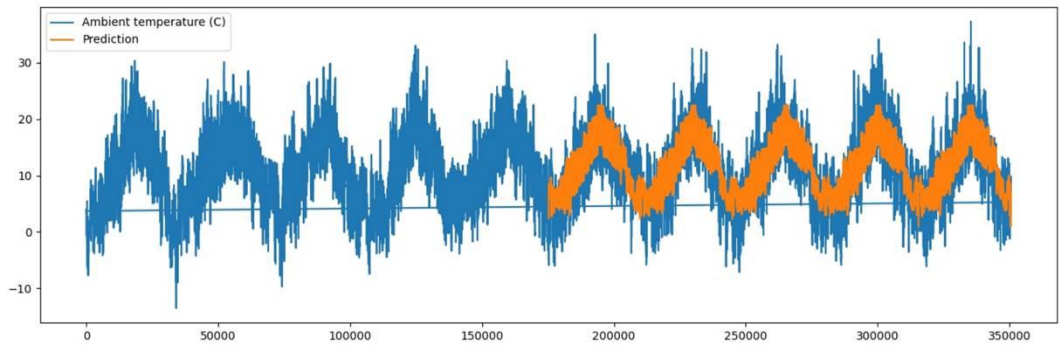

**(b)**

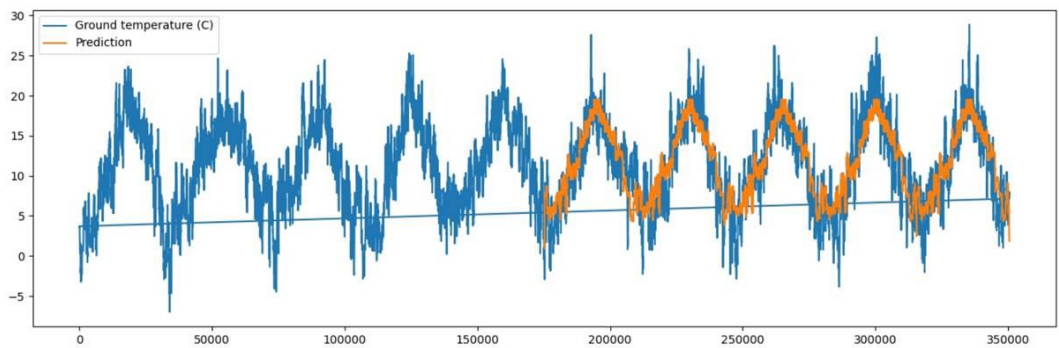

**(c)**

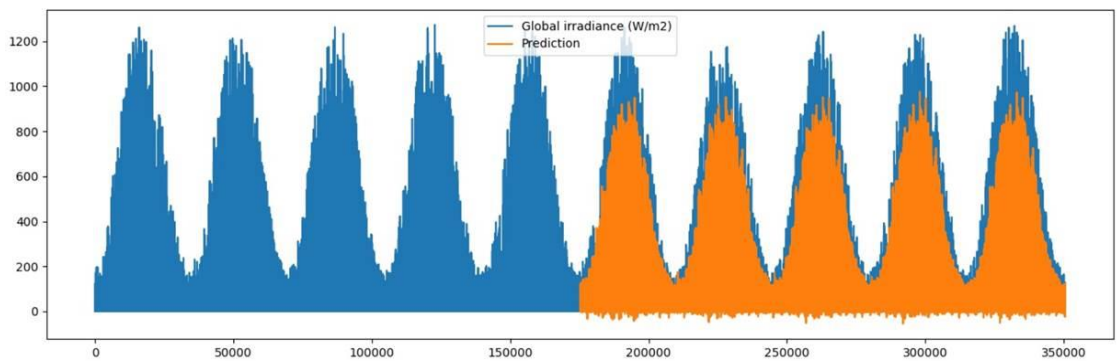

**(d)**

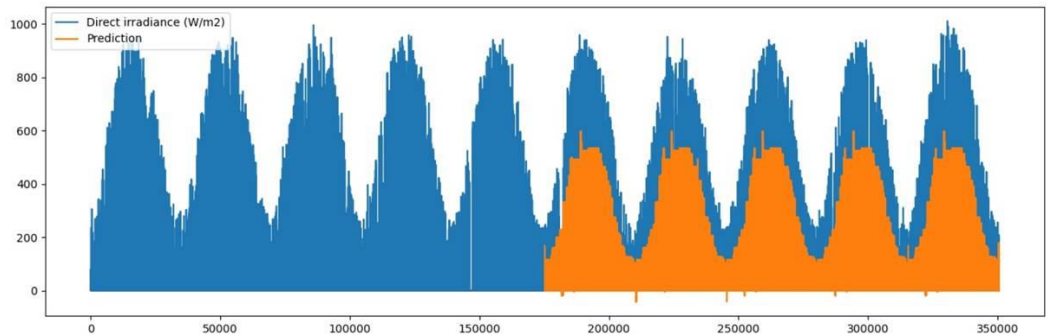

**(e)**

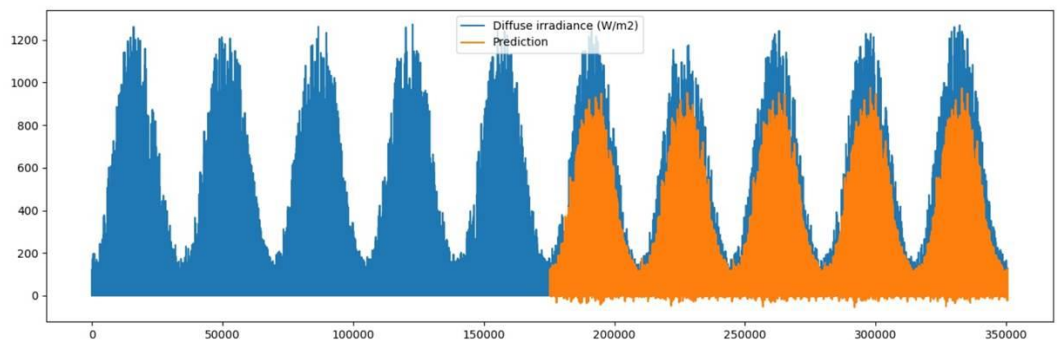

**Fig. 10:** XGboost time series model for predicting the states. (a-e) are for states 'Ambient temperature (C)','Ground temperature (C)', 'Global irradiance (W/m2)', 'Direct irradiance (W/m2)','Diffuse irradiance (W/m2)'

Next, we construct a forward problem that will map these 5 weather states with the 2 controller inputs which are *'Heat pump heat supply(kw)* and *Heat pump electrical load(kw)* , making 7 variables, to predict the internal temperature. Several regression algorithms were tested as seen in Fig.11(b). below;

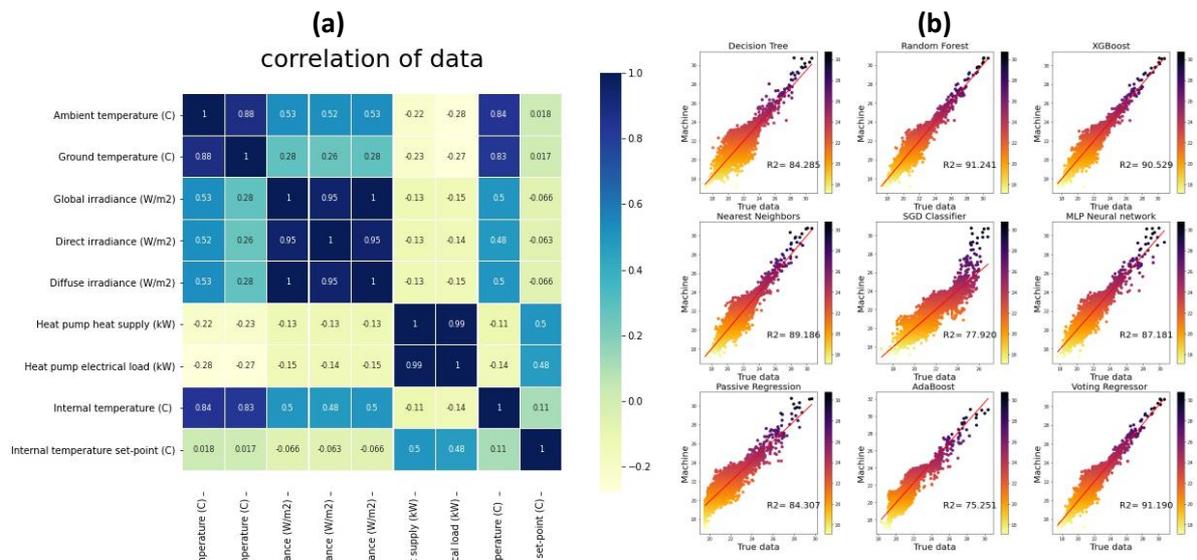

**Fig.11:** (a) Correlation map of data (b) Various regression algorithm to approximate the forward mapping of 7 to 1 on test data

In all, Random Forest showed the best $R^2(\%)$ accuracy. But Random Forest fails in optimising to the set point temperature because during optimisation, the prediction from Random Forest is based on an ensemble of decision tress, and using ensemble averaging for discrete Heaviside signals like the setpoint (17.5C /21C) gave in-accurate reconstruction as shown in Fig. 12.

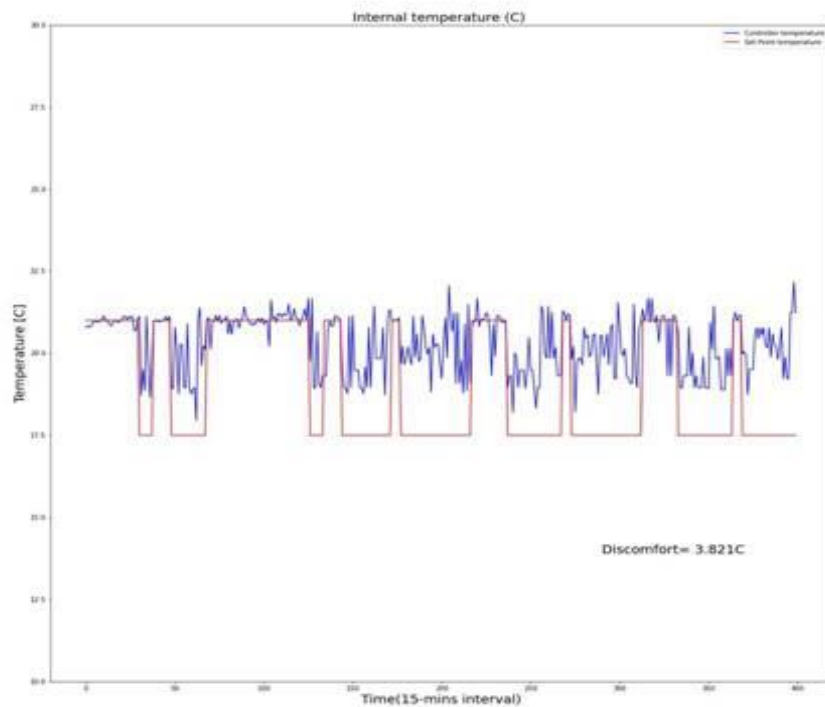

**Fig. 12:** XGboost time series model and RandomForest forward model. Discomfort is 3.821C and it is due to the ensemble averaging nature of the Random Forest Algorithm

With that in mind deterministic algorithms like Deep neural network- DNN and Polynomial regression were tested. we first trained the forward model and had the $R^2(\%)$ accuracy for them as depicted in Fig.13.

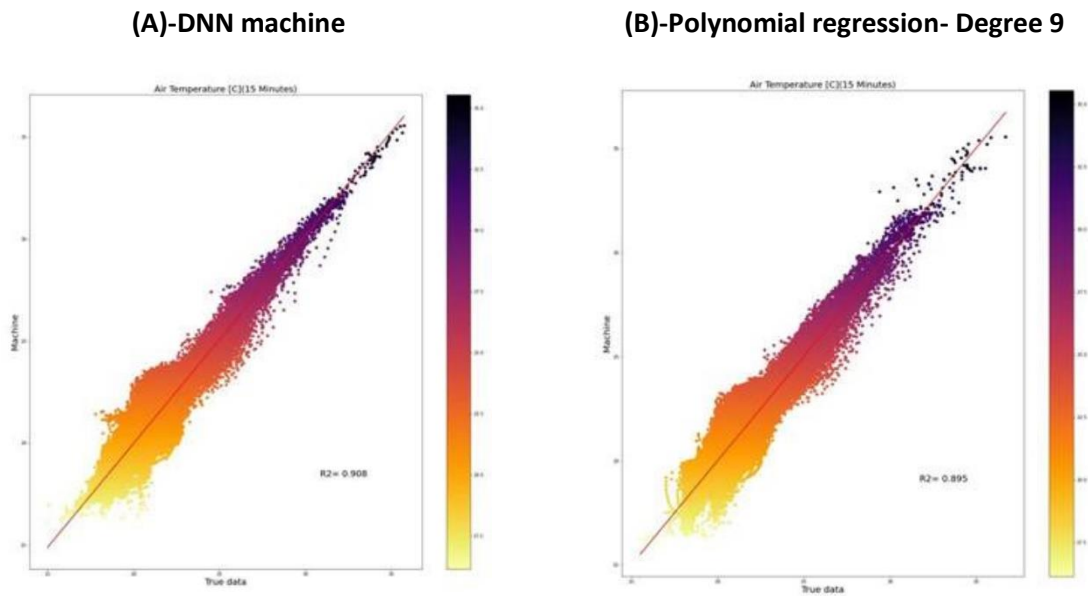

**Fig. 13:** $R^2(\%)$ accuracy on test data for the forward problem (7 to 1). (A) DNN and (B) Polynomial regression degree 9

With these in mind, the DNN controller was re-run for test set point data using XGboost as the time series machine. The discomfort for the model is depicted in Fig.(14 & 15). for DNN and Polynomial regression respectively, with Polynomial regression having a lower discomfort of **0.314C.**

Table 1 shows the summary of various combinations of XGboost series machine with 4 promising forward model, with the Polynomial regression coming 1st and DNN coming 2nd.

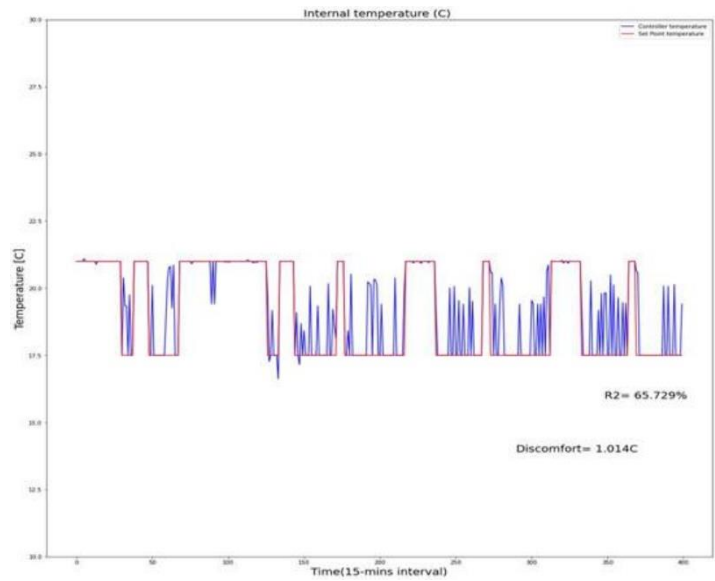

**Fig. 14:** Data driven Model- XGboost for modelling time series and DNN for modelling forward problem. Optimisation is with Nelder–Mead.

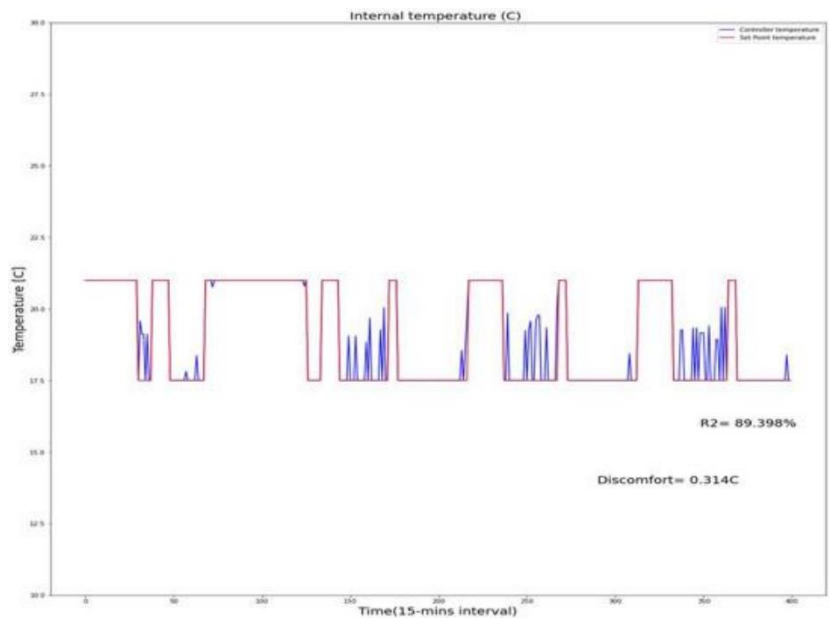

**Fig. 15:** Data driven Model-XGboost for modelling time series and Polynomial regression (Degree=9) for modelling forward problem. Optimisation is with Nelder–Mead.

| Model | Discomfort (C) |
|---|---|
| **XGboost + DNN** | 1.014 |
| **XGboost +Polynomial regression** | 0.314 |
| **XGboost +Random Forest** | 2.98 |
| **XGboost + Nearest Neighbour** | 3.26 |

**Table 1:**

we re-ran the sequence now substituting the forward model with the Cluster Classify Regress algorithm (CCR) [6 & 8] and had better performance. The prediction is for 500 steps ahead which are for 7500 minutes in the future- (6 days in advance). The forward mapping was able to trace the discontinuity of the data and better solve the inverse problem. Fig.15. are the Data-driven MPC using CCR as a forward model.

Conclusively, the approach using a CCR model as a forward model gave a better performance than one using a single machine to map the forward model.

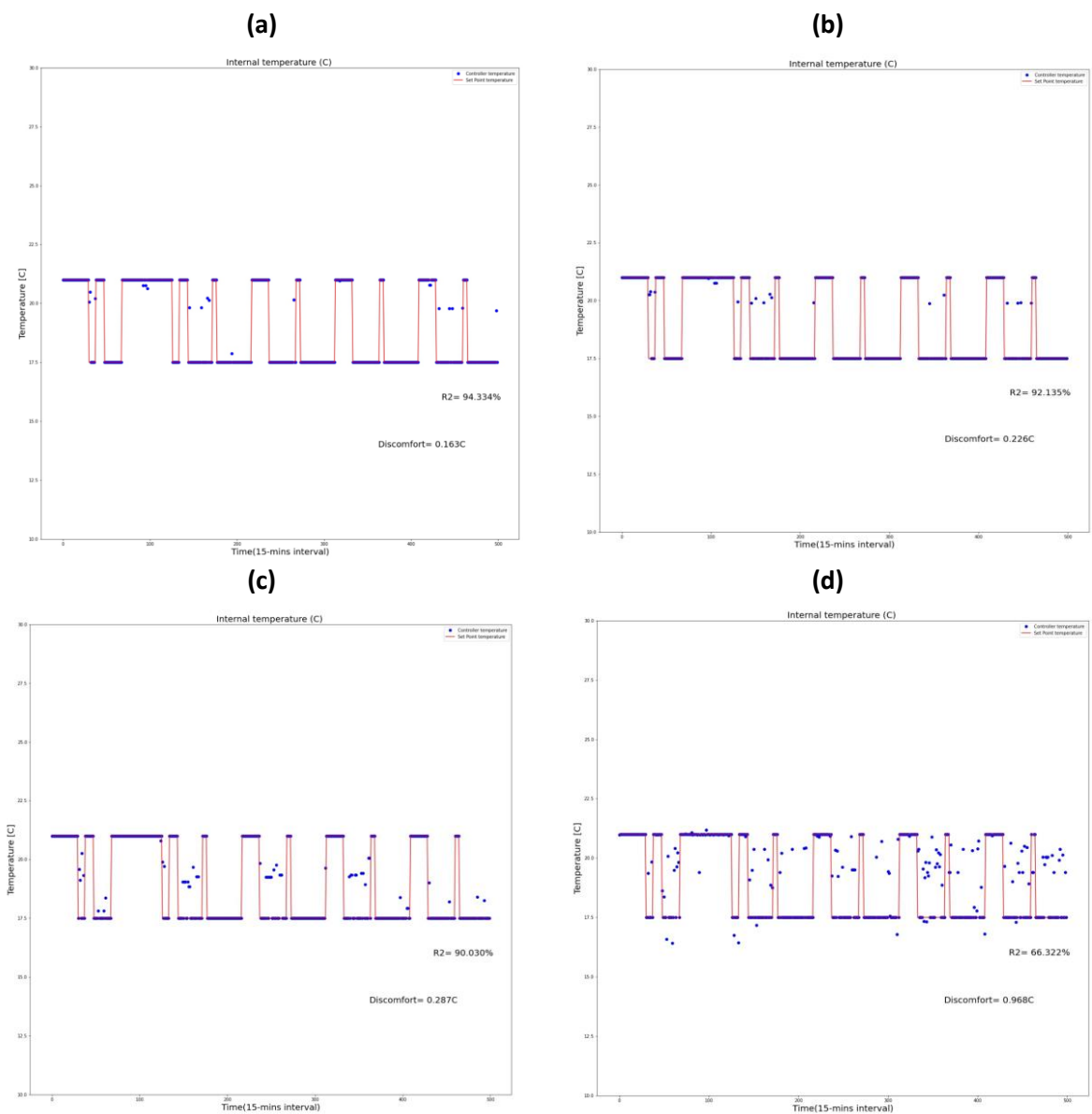

**Figure 15:** Data driven MPC Model- (a): Time series machine is an XGboost model, the forward model is a CCR model where the gate is an XGboost And the experts (2) is a 9$^{th}$-degree polynomial regressor, (d): Time series machine is an XGboost model, the forward model is a CCR model where the gate is a RandomForest and the experts (2) is a 9$^{th}$-degree polynomial regressor (c):XGboost for modelling time series and Polynomial regression (Degree=9) for modelling forward problem. Optimisation is with Nelder–Mead.(d): XGboost for modelling time series and DNN for modelling forward problem. Optimisation is with Nelder–Mead.

| Model | Discomfort (C) |
|---|---|
| (a) XGboost + CCR(XGboost/Polynomial regression) | 0.163 |
| (b) XGboost + CCR(XGboost/RandomForest) | 0.2267 |
| (c) XGboost +Polynomial Regressor | 0.28 |
| (d) XGboost + Nearest Neighbour | 0.968 |

**Table 2:**

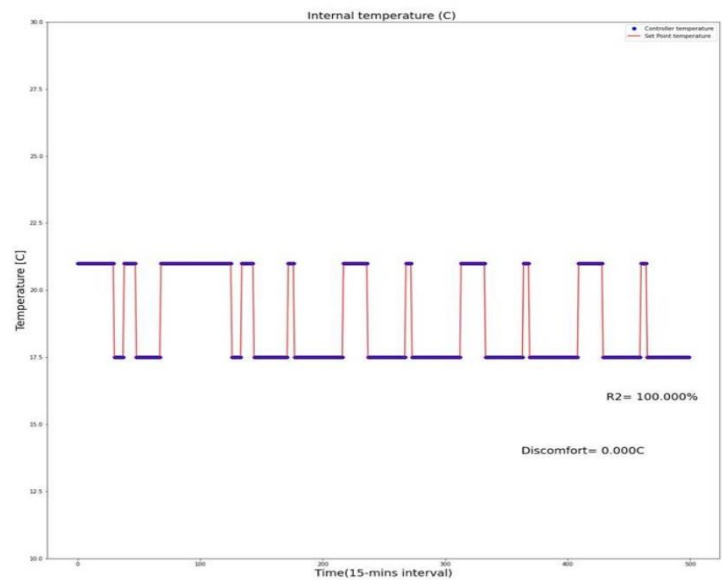

**Figure 16:**

## 5. Conclusion & Future Work

We have developed a novel model predictive controller called *CCR-MPC* for optimising the conditions of a building to track a desired set point. The approach is data driven and scales easily to any size of the data set. As with any other machine learning conjecture, the quality of the surrogate model is almost dependent on the quality of the training dataset, hence domain knowledge is important in interpreting results from this novel controller. The CCR-MPC is elegant and from the numerical results is able to track and maintain the desired set point temperature in a predictive manner. Inverse and forward UQ is also naturally imbedded from the components of the forward CCR model to the type of optimisation used in solving the inverse problem. Future work will be to apply direct reinforcement learning approach to the MPC formulation to give a natural logic of the controller learning directly from its environment and choosing the best course of action to take for future time horizons.

# 6. Acknowledgement